# A Novel Self-Attention-Enabled Weighted Ensemble-Based Convolutional Neural Network Framework for Distributed Denial of Service Attack Classification

**Kanthimathi S, Shravan Venkatraman, Jayasankar K S, Pranay Jiljith T, Jashwanth R**

School of Computer Science and Engineering, Vellore Institute of Technology, Chennai, India, 600127

**ABSTRACT** Distributed Denial of Service (DDoS) attacks are a major concern in network security, as they overwhelm systems with excessive traffic, compromise sensitive data, and disrupt network services. Accurately detecting these attacks is crucial to protecting network infrastructure. Traditional approaches, such as single Convolutional Neural Networks (CNNs) or conventional Machine Learning (ML) algorithms like Decision Trees (DTs) and Support Vector Machines (SVMs), struggle to extract the diverse features needed for precise classification, resulting in suboptimal performance. This research addresses this gap by introducing a novel approach for DDoS attack detection. The proposed method combines three distinct CNN architectures: SA-Enabled CNN with XGBoost, SA-Enabled CNN with LSTM, and SA-Enabled CNN with Random Forest. Each model extracts features at multiple scales, while self-attention mechanisms enhance feature integration and relevance. The weighted ensemble approach ensures that both prominent and subtle features contribute to the final classification, improving adaptability to evolving attack patterns and novel threats. The proposed method achieves a precision of 98.71%, an F1-score of 98.66%, a recall of 98.63%, and an accuracy of 98.69%, outperforming traditional methods and setting a new benchmark in DDoS attack detection. This innovative approach addresses critical limitations in current models and advances the state of the art in network security.

## I. INTRODUCTION

Network security is significantly threatened by DDoS attacks, targeting various online services and systems. It is a cyber-attack where genuine applications are denied from accessing network services [1]. These attacks overwhelm the targeted system with an excessive volume of traffic, effectively making it unavailable to legitimate users. DDoS attacks utilize a collection of compromised devices, often termed as zombies or bots, to overwhelm network infrastructure services with excessive traffic, aiming to disrupt normal operations [2]. Maintaining network service availability and integrity relies on detecting and mitigating these attacks. Conventional methods of detecting DDoS attacks, such as signature-based detection and rate-limiting techniques, are often ineffective against sophisticated and evolving attack strategies [3]. This is because these methods fail against evolving DDoS attacks since attackers frequently modify their tactics, making it difficult to recognize predefined patterns or maintain thresholds. These static defenses cannot adapt to dynamic and novel attack vectors, leaving systems vulnerable. DDoS Attacks continue to pose significant challenges to the security of Internet of Things (IoT) devices. There has been a formidable increase in the severity and frequency of DDoS Attacks in recent years [4]. According to Yandex and Qrator Labs, individuals were denied access to online resources of non-profit and various commercial organizations [5]. ML-based approaches offer a promising solution by enabling the detection of anomalies in network traffic patterns indicative of DDoS attacks.

There are two primary ways by which we can detect and prevent DDoS Attacks – Anomaly-based and Signature-based detection methods [6]. Signature-based detection methods can only detect previously known attacks and they cannot detect novel or unknown attacks where the signature is not previously known [7]. However, there are studies that show that signature-based detection methods like Snort are able to detect zero-day attacks but with a significantly lower detection rate compared to theoretically known attacks [8]. Anomaly-based detection method can detect new, unknown attacks by pinpointing anomalous conditions caused by an attack [9].

Studies in network intrusion detection (NID) have highlighted ML's efficacy in identifying DDoS attacks [10]. Conventional intrusion detection methods excel in processing slow-speed or small-scale data. However, their efficiency diminishes when confronted with large datasets or high-speed data streams. Therefore, there is a pressing need to develop and adopt new methodologies capable of effectively analyzing large-scale data to identify potential intrusions [11].

In this research, a novel self-attention-enabled DL framework is proposed for DDoS attack detection, and the performance is assessed through training and evaluation using the CIC-DDoS2019 [12] dataset. The proposed model is a weighted ensembling of three attention-enhanced CNN frameworks, which excels at identifying anomalies in network traffic patterns and classifying an incoming attack as DDoS or safe. It exhibits robustness against diverse attack variations due to the weighted ensembling of three unique DL classifiers, ensuring enhanced performance.

The three models used for weighted ensembling are CNN with XGBoost [13], CNN with LSTM [14], and CNN with Random Forest [15], all of which contain a self-attention (SA) mechanism at their ends. First, the individual models were assessed based on their performance to determine the relative contribution of each model to the ensemble. Then, these weights were used to perform a weighted ensembling of each model to create the overall DL classifier for DDoS attack detection.

The key contributions of the proposed work are as follows:

- The Self Attention-Enabled Weighted Ensemble Classifier introduces a novel approach to DDoS attack detection by addressing the existing research gap in ineffective feature extraction. This is achieved by combining three distinct neural network architectures—SA-Enabled CNN with XGBoost, SA-Enabled CNN with LSTM, and SA-Enabled CNN with Random Forest. Each architecture is trained individually yet simultaneously to capture multi-level features, thereby enhancing the overall effectiveness of feature extraction.
- Additionally, the proposed model incorporates a weighted ensemble mechanism that assigns weights based on the effectiveness of the features extracted by each model. The architecture that extracts the most relevant features is given a higher weight, while models that extract subtler features still receive a non-zero weight. This ensures that even the less prominent features, which might offer unique insights, are considered in the final decision-making process. This is the essence of the stacking-based meta-classifier, which leverages the strengths of all models to improve the detection of DDoS attacks.

The organization of this paper is as follows: Section II examines relevant literature; Section III explains the proposed classifier model; Section IV presents the experimental results; Section V concludes the research work; and Section VI suggests the future scope of the proposed work.

## II. RELATED WORKS

Gaur et al. [16] proposed a hybrid methodology aimed at the early detection of DDoS attacks on IoT devices, addressing the ongoing challenge of quickly diagnosing such attacks. The research focuses on enhancing the performance of ML classifiers by applying feature selection methods like Extra Trees, ANOVA, and chi-square on classifiers such as XGBoost, K-Nearest Neighbors, Decision Tree, and Random Forest. Utilizing the CICDDoS2019 dataset, the study demonstrated that combining XGBoost with ANOVA resulted in a significant feature reduction of 82.5% while achieving an accuracy of 98.34%. However, due to poor model inference capabilities, the study's reliance on a specific dataset and classifiers limits its generalizability across diverse IoT environments. Future work suggested includes tuning hyperparameters to enhance the detection accuracy further, thereby mitigating potential issues of overfitting and underfitting.

Tuan et al. [17] proposed an evaluation of various machine learning algorithms for detecting Botnet DDoS attacks, a significant and sophisticated threat to network security. The study assessed the performance of algorithms such as Unsupervised Learning (USML), Decision Tree (DT), Naïve Bayes (NB), Artificial Neural Network (ANN), and Support Vector Machine (SVM) on two prominent datasets, UNBS-NB 15 and KDD99, using metrics like accuracy, false alarm rate, and specificity. The results indicated that USML outperformed other methods in distinguishing between Botnet and normal network traffic, particularly on the KDD99 dataset. However, the study's focus on a limited number of datasets and attack types restricts its applicability to other scenarios. Future work suggests exploring additional datasets and introducing new ML algorithms based on neutrosophic theory for more targeted attack detection.

Can et al. [18] proposed a model for detecting and classifying DDoS attacks, focusing on the challenges posed by class imbalance in datasets. The research introduced a neural network-based approach that incorporates an automatic feature selection component, tailored to optimize the detection process. By applying this method to the CIC-DDoS 2019 dataset, the model significantly outperformed other machine learning-based approaches, particularly in managing imbalanced data through hinge loss and weighted loss techniques. However, the study highlighted limitations in handling rare and hierarchical labels, which led to suboptimal performance in specific attack classes like UDPLag. Additionally, discrepancies between training and testing data distributions further impacted the model's accuracy. Future work will aim to address these issues to enhance model robustness.

Sadaf et al. [19] proposed an intrusion detection method (Auto-IF) tailored for fog computing environments, where real-time network traffic classification as normal or malicious is crucial. The approach combines Isolation Forest (IF) and Autoencoder (AE) to detect intrusions, focusing on binary classification. The AE is trained to recognize normal data, and when attack data is encountered, it fails to encode it, resulting in a high reconstruction loss. IF further enhances the accuracy by refining this classification process. The method was validated using the NSL-KDD dataset, achieving an accuracy rate of 95.4%, outperforming several state-of-the-art methods. However, their feature extraction capabilities were not efficient, leading to relatively less optimal performance.

Ajeetha et al. [20] proposed a machine learning-based approach for detecting DDoS attacks, a significant threat in the field of cybersecurity. The research focuses on distinguishing between legitimate and malicious traffic flows to prevent the adverse effects of DDoS attacks, which can range from server failures to revenue loss. The authors introduced a detection system using Naive Bayes and Random Forest classifiers, which analyze traffic traces to classify network activity as normal or abnormal. Their experimental results showed that the Naive Bayes algorithm outperformed Random Forest, achieving a detection accuracy of 90.90%. However, the study's reliance on a single specific attack type limits its applicability in more diverse or evolving attack scenarios, because of which a wide range of features were not extracted effectively. Future research could explore the integration of additional datasets and advanced algorithms to enhance the detection system's robustness.

Deepa et al. [21] proposed a hybrid machine learning model to enhance the detection of DDoS attacks targeting the control plane in Software-Defined Networking (SDN) environments. The research highlights the vulnerability of the SDN controller, a critical component that manages the network, to DDoS attacks that can severely disrupt network operations. By combining Support Vector Machines (SVM) and Self-Organizing Maps (SOM), the study aimed to improve detection accuracy, and detection rate, and reduce the false alarm rate compared to traditional machine learning models. The experimental results confirmed that the hybrid model outperformed standalone algorithms in all measured metrics. However, the study's focus on the control plane limits its effectiveness across the entire SDN environment.

Future work is suggested to extend the approach by developing ensemble ML models for detecting DDoS attacks in the data plane, thereby providing a more comprehensive security solution.

Shieh et al. [22] proposed a novel DDoS attack detection framework that addresses the challenge of Open Set Recognition (OSR) in ML/DL systems, particularly in the context of evolving DDoS attack techniques. The framework integrates Incremental learning, Gaussian Mixture Model (GMM), and Bi-Directional Long Short-Term Memory (BI-LSTM) to detect both known and unknown DDoS attacks. The BI-LSTM is effective in distinguishing between legitimate and malicious traffic based on training data, while the GMM identifies novel attack patterns for further labelling and retraining. Experimental results using CIC-IDS2017 and CIC-DDoS2019 datasets demonstrated the model's ability to achieve up to 94% accuracy. However, the system's performance significantly degrades when encountering novel attacks, with recall dropping to 41.2% on certain datasets, highlighting the need for continuous updates and human intervention. Future research is suggested to enhance the framework's robustness by automating configurations and reducing dependency on traffic engineers.

Sharma et al. [23] introduced a novel approach to intrusion detection in network traffic using the Nearest Neighbour Distance Variance (NNDV) classifier, which leverages the variance of distances between objects to classify them as either malicious or normal. The study employed the KDD CUP-99 dataset to evaluate the effectiveness of NNDV and compared its predictive accuracy to the well-known K-Nearest Neighbour (KNN) classifier. The results demonstrated that NNDV often performs better than KNN, particularly with normalized data, achieving accuracy rates up to 99%. The paper also explored different cross-validation techniques to further refine the model's performance. The conclusion emphasizes that while NNDV shows promise as an intrusion detection method, fine-tuning its parameters (k, p, and threshold μ) could enhance accuracy even further. Unlike deep learning or neural network models, NNDV's results remain consistent across runs. The author suggests that future work should focus on optimizing these parameters, possibly through automated methods, and exploring feature selection and multiclass prediction extensions for the algorithm.

Assis et al. [24] proposed a security system for SDN environments in IoT networks, addressing the significant challenge of mitigating DDoS attacks. The authors introduced a Convolutional Neural Network (CNN) to detect such attacks in near real-time, demonstrating its superiority over other anomaly detection methods like Logistic Regression and Multi-Layered Perceptron. Their approach also includes a Game Theory-based mitigation strategy to optimize packet discard rates, enhancing SDN resilience.

Although effective, the system's performance gets constrained by the complexity of larger, more varied SDN environments and potential limitations in handling evolving attack strategies.

Gohil et al. [25] explored the detection of DDoS attacks, focusing on the effectiveness of various supervised classification algorithms. Their research aimed to accurately distinguish between malicious and legitimate network traffic using the CICDDoS2019 dataset, which includes recent DDoS attack signatures. The study found that tree-based classifiers like Decision Tree and Random Forest, as well as distance-based classifiers like K-Nearest Neighbors (K-NN), outperformed other models in accuracy. However, the research is limited by the scope of the dataset and the types of DDoS attacks considered, indicating a need for further expansion and testing on a broader range of attacks in future work.

Rajagopal et al. [26] proposed a meta-classification approach for network intrusion detection, focusing on both binary and multiclass classification, and implemented it on the Azure cloud platform. The study aimed to address the evolving threat landscape by optimizing hyperparameters and feature subsets to improve detection accuracy. Using robust classifiers like Decision Jungle, Bayes Point Machine, and Logistic Regression, the model was tested on datasets including UNSW NB-15, CICIDS 2017, and CICDDOS 2019, achieving high accuracy rates. Although the model demonstrated strong performance, its reliance on predefined datasets limits its adaptability to novel attacks. Future research should explore incorporating additional datasets and employing deep learning within cloud-based environments.

Niyaz et al. [27] addressed the challenge of developing an efficient and flexible Network Intrusion Detection System (NIDS) capable of handling unforeseen network security breaches. They proposed a DL-based approach using self-taught learning (STL), which was applied to the NSL-KDD dataset as a benchmark for network intrusion detection. The research demonstrated that their approach outperformed previous NIDS implementations in terms of accuracy, precision, recall, and f-measure. Despite its promising results, the study's limitations include the potential need for more advanced techniques like Stacked Autoencoders and other classifiers for further improvement. Future work should focus on real-time implementation and feature learning from raw network traffic.

Perakovic et al. [28] proposed a specialized DDoS detection model to address the increasing challenge of mitigating DDoS attacks in smart home environments, which often consist of a diverse array of IoT devices. The research tackles the problem by categorizing IoT devices into four distinct classes based on their traffic predictability and applying a boosting method of logistic model trees tailored to each class. This innovative approach departs from traditional models that assume a single legitimate traffic profile for all devices, thereby improving detection accuracy, which the study reports to be between 99.92% and 99.99% for the different classes. The study's findings demonstrate the model's high efficacy, but it also highlights a potential limitation: the reliance on predefined device classes may restrict the model's adaptability to new and evolving IoT devices. Consequently, the authors suggest further exploration into extending this approach to other IoT environments and attack types to enhance its generalizability and robustness.

Singh et al. [29] conducted a survey on Distributed Denial-of-Service (DDoS) attacks, focusing on the vulnerabilities of web-enabled platforms like IoT, cloud computing, and Software-Defined Networking (SDN). The authors underscore the distributed nature of these platforms, which has made them increasingly susceptible to large-scale DDoS attacks. They review a variety of detection methods, including the K-Nearest Neighbors (KNN)-based DDoS detection algorithm by Dong & Sarem (2019), which shows potential but needs further refinement for real-time application in SDN environments. Another significant work cited is Yuan et al.'s (2017) DeepDefense system, which employs Recurrent Neural Networks (RNNs) and Convolutional Neural Networks (CNNs) to analyze historical packet data, though it is highly resource-intensive. The survey also references the LUCID technique by Doriguzzi-Corin et al. (2020), a lightweight CNN-based method aimed at distinguishing between malicious and safe traffic, though it still faces challenges in scalability. Singh et al. conclude that while existing solutions show promise in controlled environments, they are often hampered by high computational costs and real-world implementation difficulties, particularly in resource-constrained settings like IoT networks. This highlights the need for more efficient, adaptable, and scalable defenses against the evolving nature of DDoS attacks.

Anupama Mishra et al. [30] proposed a defensive mechanism to detect and mitigate DDoS attacks in Software-Defined Networks (SDNs) integrated with cloud computing. The research addresses the challenge of detecting DDoS attacks, which remain a significant threat despite existing solutions. Their approach leverages variations in entropy and flow table attributes to recognize and alleviate DDoS attacks with low computational overhead. The proposed method demonstrates high detection rates (98.2%) and a low false positive rate (0.04%) under different attack strengths. While the research shows promising results, it is limited by its focus on a single controller and does not address slow DDoS attacks that mimic legitimate traffic. Future work could explore the use of multiple controllers and extend detection to more subtle attack types.

Kathirkamanathan et al. [31] proposed a solution to mitigate Distributed Denial of Service (DDoS) attacks targeting financial services, which are particularly vulnerable in cloud computing environments. The study focused on using deep neural networks, specifically stacked Long Short-Term Memory (LSTM) networks, to classify and mitigate DDoS attacks at the HTTP layer of transactional endpoints. The authors compared various neural networks, including simple Artificial Neural Networks (ANN) and Gated Recurrent Units (GRU), concluding that stacked LSTM outperformed the others, achieving an accuracy of 99.3%. However, the research was limited to HTTP-based flooding attacks, with no consideration for other types of DDoS attacks like UDP or ICMP. Future work could involve extending the model to handle other attack types and automating the tuning of hyperparameters for improved performance.

Najafimehr et al. [32] proposed a hybrid machine learning method to address the growing threat of DDoS attacks on network service availability. The problem they identified was that traditional machine learning algorithms, though effective against known DDoS attacks, struggle to detect previously unseen attacks. Their contribution combines an unsupervised clustering technique, DBSCAN, with a supervised classification model, allowing for the separation of anomalous traffic and the labeling of clusters based on statistical measures. The method demonstrated significant improvements in detection performance when tested on CICDDoS2019 data. While the approach shows strong results, a possible limitation is the reliance on accurate feature engineering and proper parameter tuning, which may affect scalability across different network environments.

Batham et al. [33] proposed a CNN DL technique for detecting botnet attacks within IoT applications. The problem addressed in the study is the increasing threat of botnet attacks, which involve devices infected with malware and remotely controlled to carry out large-scale cyber assaults. To combat this, the research leverages machine learning and deep learning classification methods, particularly focusing on CNN and LSTM networks. The study used the Bot-IoT dataset to validate the effectiveness of the proposed method in both binary and multi-class classification scenarios. The results demonstrated a significant improvement in detection accuracy, achieving 99.80%, with a reduced error rate of 0.20%. However, the method still has limitations in terms of generalization across different types of botnet attacks or IoT devices not represented in the dataset, along with suboptimal feature extraction. Further research is necessary to test its robustness in more diverse environments.

The summary of the literature survey has been presented in Table 1.

**Table 1: Summary of Literature Review**

| Author(s) | Problem Statement | Research Contribution | Limitations |
|---|---|---|---|
| Gaur et al. [16] | Timely identification of DDoS attacks on IoT devices. | Hybrid methodology using feature selection (Extra Trees, ANOVA, and chi-square) with classifiers (RF, DT, k-NN, XGBoost). | Reliance on specific datasets and classifiers limits generalizability. |
| Tuan et al. [17] | Detection of Botnet DDoS attacks in network security. | Evaluation of ML algorithms (SVM, ANN, NB, DT, USML) on datasets (KDD99, UNBS-NB 15) using false alarm rate, accuracy, and specificity. | Focus on specific datasets and attack types may restrict applicability to other scenarios. |
| Can et al. [18] | Detection and classification of DDoS attacks, addressing class imbalance. | Neural network-based approach with automatic feature selection and weighted loss techniques. | Limitations in handling rare and hierarchical labels; discrepancies in data distributions. |
| Sadaf et al. [19] | Intrusion detection in fog computing environments. | Auto-IF method combining Autoencoder (AE) and Isolation Forest (IF) for binary classification. | Limited to binary classification, potentially overlooking more nuanced attack types. |

| Ajeetha et al. [20] | Detection of DDoS attacks to prevent server failures and revenue loss. | Detection system using Naive Bayes and Random Forest classifiers to classify network activity. | Reliance on a single dataset and a specific attack type limits applicability. |
|---|---|---|---|
| Deepa et al. [21] | Enhancing DDoS attack detection in SDN environments targeting the control plane. | Hybrid model combining SVMs and SOMs. | Focus on the control plane may limit effectiveness across the entire SDN environment. |
| Shieh et al. [22] | DDoS attack detection in evolving attack scenarios using Open Set Recognition (OSR). | Framework integrating BI-LSTM, GMM, and incremental learning for detecting known and unknown attacks. | Performance degrades when encountering novel attacks, with a significant drop in recall. |
| Sharma et al. [23] | Intrusion detection in network traffic using the Nearest Neighbor Distance Variance (NNDV) classifier. | NNDV classifier applied to the KDD CUP-99 dataset, compared with KNN classifier. | Requires fine-tuning of parameters (k, p, and threshold $\mu$) for optimal performance. |
| Marcos et al. [24] | Security system for SDN environments in IoT networks to mitigate DDoS attacks. | Introduced CNN for detection and a Game Theory-based mitigation strategy. | Potential limitations in handling evolving attack strategies and the complexity of larger, more varied SDN environments. |
| Gohil et al. [25] | Detection of DDoS attacks using supervised classification algorithms. | Evaluation of various classifiers on the CICDDoS2019 dataset. | Research is limited by the scope of the dataset and the types of DDoS attacks considered. |
| Niyaz et al. [26] | Development of efficient Network Intrusion Detection System (NIDS). | DL-based approach using Self-taught Learning (STL) applied to NSL-KDD dataset. | Reliance on a benchmark dataset may not fully represent real-world scenarios. |
| Rajagopal et al. [27] | Meta-classification approach for network intrusion detection in cloud environments. | Optimization of hyperparameters and feature subsets on datasets like UNSW NB-15, CICIDS 2017. | Reliance on predefined datasets may limit adaptability to novel attacks. |
| Perakovic et al. [28] | DDoS detection in smart home environments with diverse IoT devices. | Categorization of IoT devices and application of logistic model trees. | Reliance on predefined device classes may restrict adaptability to new and evolving IoT devices. |
| Singh et al. [29] | Survey on DDoS attacks in web-enabled platforms like IoT, cloud computing, and SDN. | Review of detection methods, including KNN-based and DeepDefense (RNNs and CNNs) systems. | Often hampered by high computational costs and real-world implementation difficulties, particularly in resource-constrained settings. |
| Mishra et al. [30] | Detection and mitigation of DDoS attacks in SDNs integrated with cloud computing. | Entropy variation and flow table attributes for DDoS detection. | Focus on a single controller; does not address slow DDoS attacks mimicking legitimate traffic. |

| Kathirkamanathan et al. [31] | DDoS attack mitigation in financial services using deep neural networks. | Stacked LSTM networks for classifying and mitigating HTTP-based DDoS attacks. | Limited to HTTP-based flooding attacks, no consideration for other types of DDoS attacks like ICMP or UDP. |
| Mohammad Najafimehr et al. [32] | DDoS attacks are a growing threat to service availability, and traditional ML techniques struggle to detect unknown attacks. | Developed a hybrid ML method combining DBSCAN clustering (unsupervised) with classification algorithms (supervised) to detect unknown DDoS traffic. Tested the method on real datasets (CICIDS2017 and CICDDoS2019). | The effectiveness of the method may heavily rely on feature engineering and parameter tuning, which could limit adaptability to various network environments and attack types. |
| Batham et al. [33] | Detection of botnet attacks in IoT applications using deep learning. | CNN and LSTM networks applied to the Bot-IoT dataset for classification. | Method may have limitations in generalizing across different types of botnet attacks or IoT devices not represented in the dataset. |

From the table 1, the following research gaps have been identified:

- **Limited Generalization of Feature Selection Techniques:** Several studies utilized traditional feature selection techniques such as Extra Trees, ANOVA, and chi-square. While these methods have shown promising results for specific datasets, they may not generalize well to other diverse or evolving datasets, particularly in real-time environments. The challenge lies in developing more adaptive feature extraction methods that can dynamically learn relevant features in varying conditions, without relying on predefined selections.

- **Limited Adaptability of Traditional Machine Learning Models:** Numerous employed traditional ML models such as Random Forest, Naive Bayes, and Decision Trees. While effective for specific attack types or datasets, these models exhibit limited adaptability to evolving attack patterns and cannot handle the complexity of modern network environments. There is a need for models that can learn continuously and evolve with emerging attack vectors.

- **Challenges in Handling Novel Attacks:** Approaches such as Open Set Recognition attempt to address novel attack detection. However, the performance degrades when faced with unseen attacks. Existing models focus heavily on known attack types, while the ability to detect novel, unknown threats remains limited. A more robust approach to model learning is needed that can adapt to new threats without prior knowledge, ensuring effective defense against zero-day attacks.

## III. PROPOSED WORK

This section presents a novel Self-Attention (SA) Enabled Weighted Ensemble Classifier for detecting DDoS attacks, combining 3 different individual classification algorithms within the overall ensemble classifier. Initially, the CIC-DDoS2019 benchmark dataset undergoes extensive data preprocessing, including merging individual datasets, managing high feature correlations, handling null values statistically, and selecting the most relevant features. After preprocessing, the preprocessed data is split into training and testing subsets, with 80% allocated for training the proposed model and 20% reserved for its validation. This allocation is made to ensure a balance between having enough data for training and ensuring a reliable testing process. Each model, a

Self-Attention-enabled Convolutional Neural Network (CNN) combined with XGBoost, LSTM, and Random Forest, is designed to enhance feature representation and improve the detection of DDoS attacks.

The trained models are then evaluated, and their performance results are used to determine their contribution weights in a weighted ensemble approach. This ensemble combines the strengths of each individual model through a stacking-based weighted prediction mechanism, improving the overall accuracy of the system. Finally, the model's effectiveness is assessed using key metrics such precision, accuracy, F1 score, and recall, ensuring a robust and reliable detection system for various types of DDoS attacks. The flow of work has been demonstrated in Figure 1.

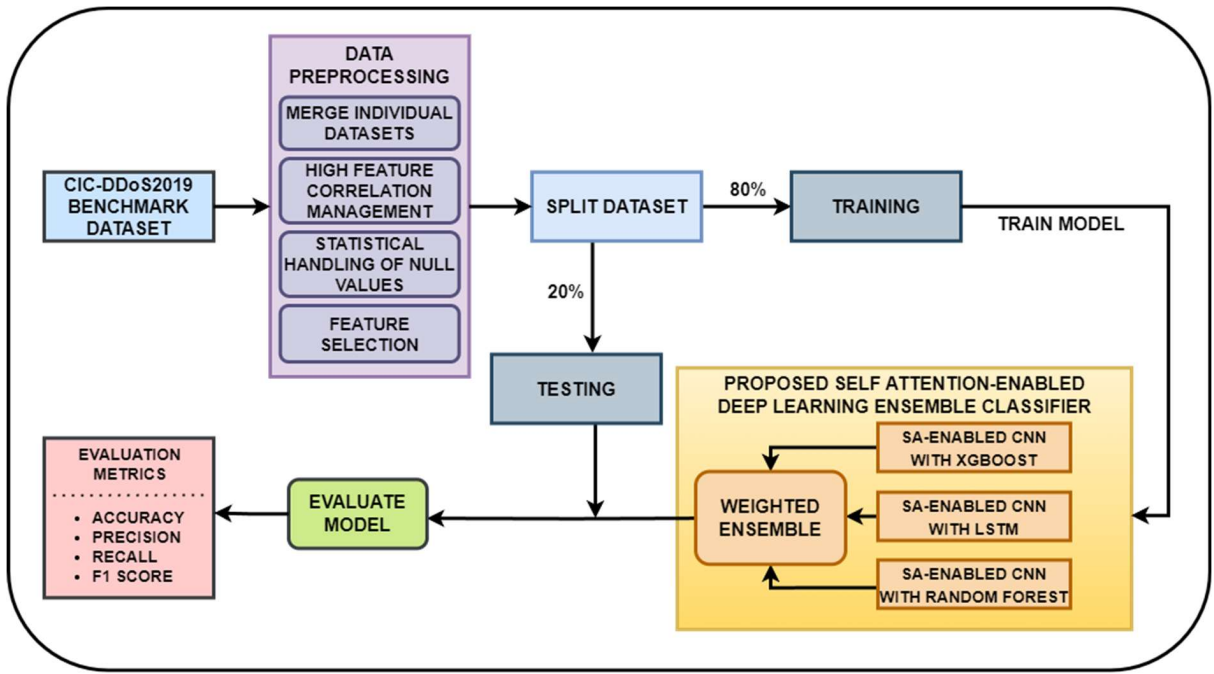

Figure 1: Workflow of Proposed Model Framework

## A. DATASET DESCRIPTION AND PREPROCESSING

This research is based on the CIC-DDoS2019 dataset, which comprises real-world network traffic data to evaluate DDoS attack detection algorithms. The dataset offers diverse DDoS attack scenarios and realistic traffic patterns, aiding in developing robust detection models. With comprehensive labelling and widespread use in research, it serves as a trusted benchmark for evaluating DDoS detection techniques. The total dataset is formed by merging eight different attack files such as PortMap, NetBIOS, and LDAP. Preprocessing involved the elimination of duplicate rows, the management of highly correlated columns, and the statistical handling of null values. Feature selection involved carefully choosing columns that provide essential information for identifying DDoS attacks. Among the selected columns are source and destination IPs, which are fundamental for identifying the origin and destination of network traffic. By analyzing these IPs, patterns, and anomalies indicative of DDoS attacks can be detected. Additionally, source and destination ports are included to gain insights into the specific services or applications being targeted by potential attackers. Considering features like protocols and timestamps provides valuable contextual information for understanding the nature and timing of network activities.

The 'Source Port' and 'Destination Port' columns indicate the source and destination ports of network traffic, providing insights into the specific services or applications involved in communication. The 'Protocol' column specifies the protocol used for communication, such as ICMP, TCP, or UDP, aiding in understanding the communication mechanism between network entities, while the term 'Total Length of Fwd Packets' represents the cumulative length of packets forwarded within a flow, offering information on the amount of data transmitted. The 'Flow IAT Mean' represents the average time between two consecutive packets in a flow, aiding in analyzing traffic patterns' regularity or irregularity, and the 'Fwd Packets/s' feature calculates the rate of forwarded packets per second, providing insights into traffic intensity. The 'Packet Length Mean' represents the average packet length in a flow, useful for detecting anomalies or patterns in packet sizes. To ensure quality and relevance of the features, a heatmap was generated to visualize the correlation between features, helping to identify and eliminate highly correlated variables that could potentially introduce multicollinearity into the model. This correlation matrix, as illustrated in Figure 2, provides a comprehensive overview of the relationships between all numerical features.

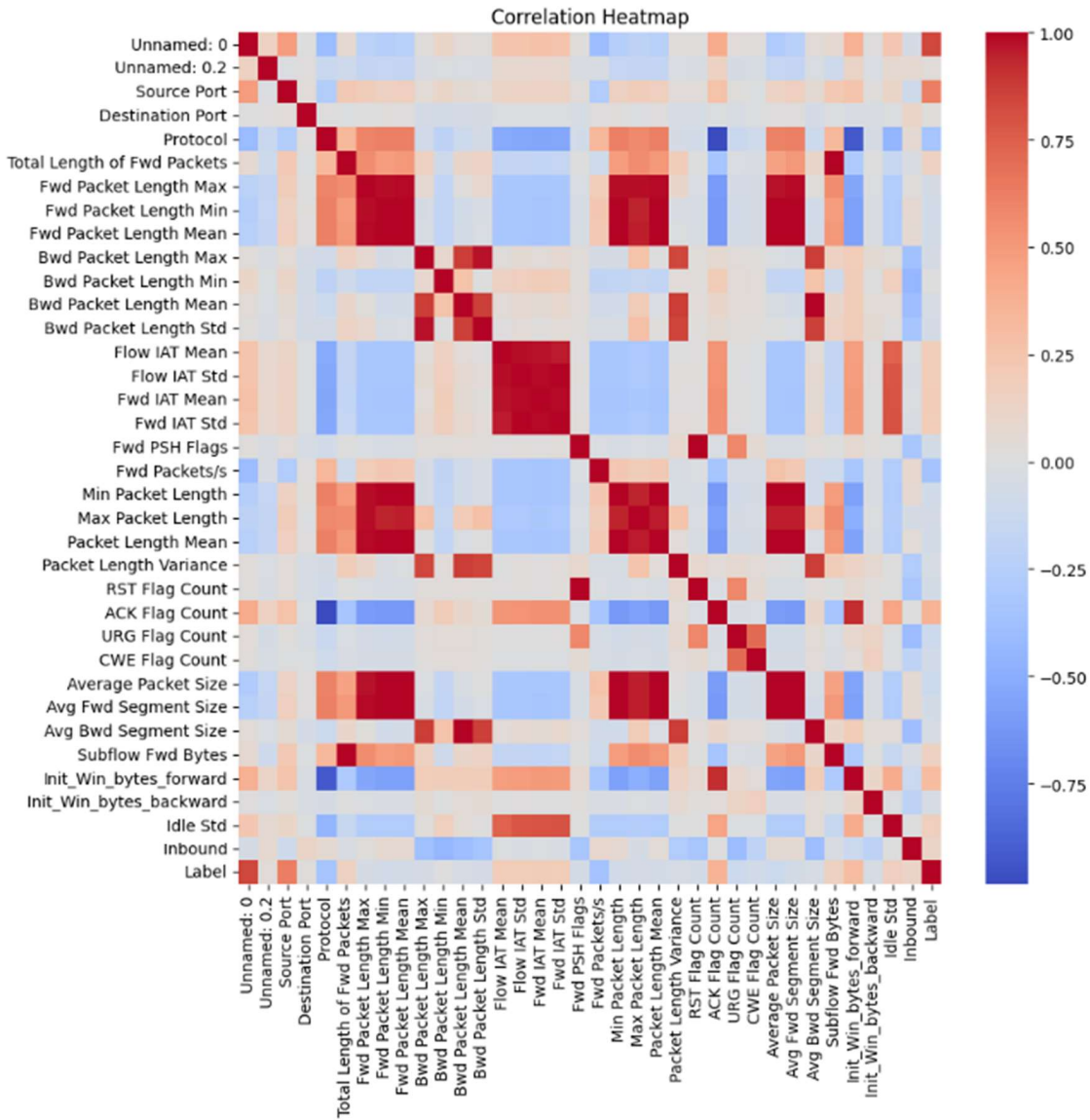

Figure 2: Heatmap Visualizing the Correlation Between Features

Furthermore, a Pareto chart was employed to highlight the most significant factors contributing to the variance in the dataset. It highlights the distribution of various features within the dataset, demonstrating that a small subset of values for each feature accounts for the majority of occurrences. For instance, in features like Source Port and Protocol, a few specific values dominate the dataset, contributing to a steep rise in the cumulative percentage. This indicates that certain values play a disproportionately significant role, which aligns with the Pareto principle (80/20 rule). This suggests that focusing on these dominant values could yield the most significant impact in detecting patterns or anomalies. The Pareto analysis results are depicted in Figure 3.

The features discussed above are chosen during feature extraction and following this, all features except the target column were scaled to normalize their range between [0, 1]. This preprocessed data consisting of 1,341,858 rows, is then used for training the proposed model.

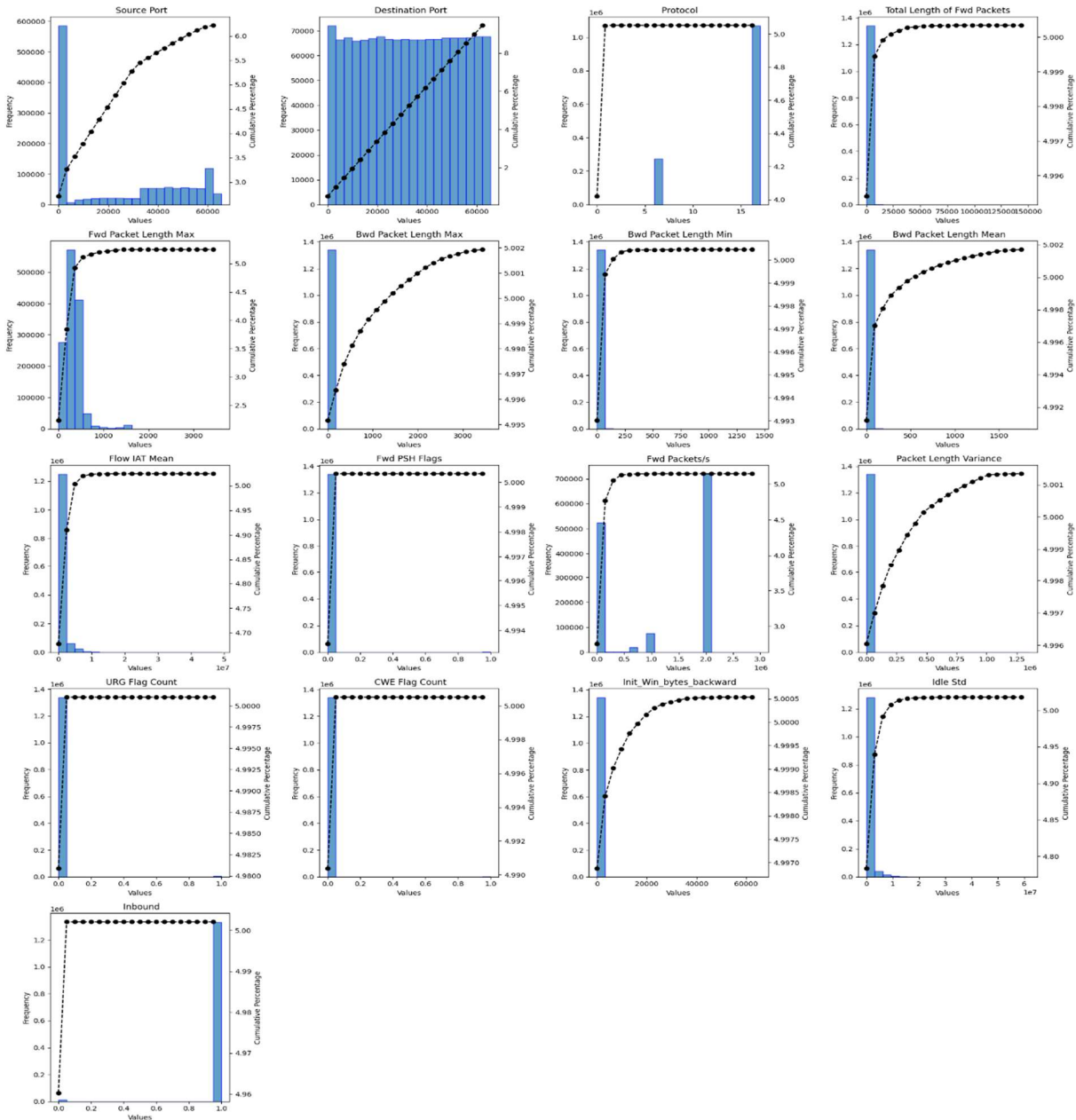

Figure 3. Pareto Chart of Key Features Contributing to Variance

## B. PROPOSED WEIGHTED ENSEMBLE CLASSIFIER

The proposed model is a weighted ensemble of three deep neural networks – SA-enabled CNN with XGBoost, SA-enabled CNN with LSTM, and SA-enabled CNN with Random Forest. The core strength of the proposed model lies in its unique diverse feature extraction capabilities. Each of the three neural networks receives preprocessed data individually, allowing them to extract different intricate features. Additionally, a self-attention mechanism is incorporated at the end of each neural network to further refine the feature extraction process. This emphasizes additional focus on the most critical features for training, thereby improving its predictive capabilities. Finally, predictions are obtained from individual models combined using a stacking-based weighted sum approach to collectively include inferences from each model. This process ensures that the model comprehensively learns how different features impact its predictions. The combined inferences are then processed through a fully connected layer, followed by a sigmoid activation function, to classify the input as either a DDoS attack or safe. Existing models rely on a single CNN or traditional ML algorithms such as SVMs and decision trees for feature extraction and fail to effectively capture the diverse range of features required for precise classification, resulting in suboptimal performance. The proposed model tackles the challenge of limited feature extraction by using three CNN backbones to capture features at different scales. A self-attention mechanism at each stage refines these features, resulting in a more comprehensive understanding of the data. This diversity in feature extraction is crucial for capturing a broad spectrum of relevant details that improve prediction accuracy.

### 1) SELF-ATTENTION ENABLED CNN WITH XGBOOST

The preprocessed input undergoes convolutional operations (1) in the first layer. Here, 128 filters with a kernel size of 3 convolve across the input, extracting localized patterns and DDoS-relevant features. Following this, batch normalization (2) and ReLU activation (3) functions are applied to stabilize and introduce non-linearity to the convolved output. These features are then fed to another set of convolutions, batch normalization, and activation layers. The second convolutional layer applies an additional 64 filters, allowing the model to capture higher-order features essential for distinguishing subtle differences between DDoS attacks and safe traffic. The batch normalization layer then normalizes these new feature maps, and the ReLU activation enables learning of non-linear decision boundaries. The features then undergo self-attention (SA) (4), where they are transformed into query (Q), key (K), and value (V) vectors. The SA mechanism (SAM) calculates the query and key vectors' dot product to obtain attention scores, which are scaled to ensure gradient stability and prevent large values. Then, they are normalized via softmax (5) to indicate feature importance and fed into an XGBoost classifier, where sequential decision trees are trained to produce the predicted probabilities for DDoS attacks or safe traffic.

$$O(m, n, o) = \sum_{x=0}^{M-1} \sum_{y=0}^{N-1} \sum_{z=0}^{C-1} I(m + x, n + y, z) \cdot K(x, y, z, o) + b \quad (1)$$

where,

$O(m, n, o) \leftarrow$ output feature map at position $(m, n)$

$I(m + x, n + y, z) \leftarrow$ input feature map at position $(m + x, n + y)$

$K(x, y, z, o) \leftarrow$ convolutional kernel

$b \leftarrow$ bias

$m \leftarrow$ width of convolutional kernel

$n \leftarrow$ height of convolutional kernel

$c \leftarrow$ number of input channels

$$y_i = \gamma \left( \frac{x_i - \mu_B}{\sqrt{(\sigma_B^2 + \varepsilon)}} \right) + \beta \quad (2)$$

$$y(x) = (0, x) \quad (3)$$

$$SA(Q, K, V) = softmax \left( \frac{QK^T}{\sqrt{d_k}} \right) V \quad (4)$$

$$softmax(p_x) = \frac{e^{Px}}{\sum_{y=1}^{N} e^{Py}} \quad (5)$$

### 2) SELF-ATTENTION ENABLED CNN WITH LSTM

The preprocessed data is convolved (1) using 128 filters to extract indispensable spatial features critical for detecting DDoS attacks. Following this, long short-term memory (LSTM) (6-10) layers process the sequential data, capturing temporal dependencies and long-term interactions within the input data. With its ability to retain information over time, LSTM units effectively analyze the dynamic patterns inherent in network traffic, allowing for the identification of subtle changes and anomalies indicative of DDoS attacks. ReLU activation (3) adds non-linearity to enhance feature representations. In the subsequent repeat, the process iterates through convolutional, LSTM, and ReLU layers once again, refining the model's understanding of the input data and capturing increasingly nuanced patterns and relationships. SAM (4) is applied to the refined features, dynamically adjusting the model's focus to highlight crucial aspects of the data. The highlighted features are pooled using an average pooling layer to reduce the spatial dimensions of the features while retaining their essential information. Finally, a sigmoid activation function is applied to the pooled features, producing a probability score for each feature map.

Input Gate,

$$i_t = \sigma(W_{xi} x_t + W_{hi} h_{t-1} + W_{ci} c_{t-1} + b_i) \quad (6)$$

Forget Gate,

$$f_t = \sigma(W_{xf} x_t + W_{hf} h_{t-1} + W_{cf} c_{t-1} + b_f) \quad (7)$$

Cell State,

$$c_t = f_t c_{t-1} + i_t \tanh \tanh (W_{xc} x_t + W_{hc} h_{t-1} + b_c) \quad (8)$$

Output Gate,

$$o_t = \sigma(W_{xo} x_t + W_{ho} h_{t-1} + W_{co} c_{t-1} + b_o) \quad (9)$$

Hidden State (output),

$$h_t = o_t \odot \tanh (c_t) \quad (10)$$

### 3) SELF-ATTENTION ENABLED CNN WITH RANDOM FOREST

The preprocessed input is convoluted (1) with 128 filters to extract localized DDoS-relevant features. Batch normalization (2) and ReLU activation (3) are applied to mitigate internal covariate shift, after which a second convolutional layer with 64 filters captures higher-order features. This is followed by another round of batch normalization and ReLU activation to normalize the features and introduce non-linearity. These refined features are fed into a SAM (4), transforming them into value (V), key (K), and query (Q) vectors to compute scaled attention scores, which are normalized using a softmax function. The SA outputs are then input into a Random Forest classifier, where multiple decision trees classify the data as either a DDoS attack or safe, with final predictions made by aggregating the individual tree results.

### 4) STACKING-BASED WEIGHTED ENSEMBLE CLASSIFIER

A stacking-based weighted sum methodology is applied to integrate the output probabilities from each classifier, aiming to consolidate their performances and create a more resilient and high-performing model. Stacking, also known as stacked generalization, is an ensemble technique aimed at mitigating overfitting by allowing the global classifier to generalize from the predictions of the base classifiers. Different base classifiers - SA-Enabled CNN with XGBoost (3.2.1), SA-Enabled CNN with LSTM (3.2.2), and SA-Enabled CNN with Random Forest (3.2.3) – perform well in extracting different scales and contexts of features, resulting in predictions framed out of different learning algorithms. A weighted sum was computed to integrate the obtained predictions from these models, labelled as α, β, and γ. The weights for α, β, and γ - 0.34, 0.41, and 0.25, respectively - are determined based on the model's performances (as depicted in Table 1) using a grid-search algorithm, which is presented in Algorithm 1.

| **Algorithm 1: Grid Search for Ensembling Weights** |
| --- |

$models \leftarrow [model_1, model_2, model_3]$

$y_{true} \leftarrow true\ values$

$df \leftarrow dataframe\ to\ store\ weights\ for\ final\ calculation;$

$for\ w_1 \leftarrow 0\ to\ 4\ do$

    $for\ w_2 \leftarrow 0\ to\ 4\ do$

        $for\ w_3 \leftarrow 0\ to\ 4\ do$

            $wts \leftarrow [\frac{w_1}{10}, \frac{w_2}{10}, \frac{w_3}{10}]$

            $preds_{weighted} \leftarrow models \cdot preds\,, along\ axis: ((0),(0))$

            $ensemblePreds_{weighted} \leftarrow argmax(preds_{weighted}, axis = 1)$

            $accuracy_{weighted} \leftarrow accuracy(ensemblePreds_{weighted}, y_{true})$

            $df \leftarrow df.append(w_1, w_2, w_3, accuracy_{weighted})$

        $end\ for$

    $end\ for$

$end\ for$

$row_{maxAccuracy} \leftarrow locate(max\,(df['accuracy']))$

$\alpha \leftarrow row_{maxAccuracy}[0]$

$\beta \leftarrow row_{maxAccuracy}[1]$

$\gamma \leftarrow row_{maxAccuracy}[2]$

$\mathbf{return}\ \alpha, \beta, \gamma$

$\mathbf{end}$

These weighted features are fed as input to the global classifier after stacking, which is followed by a fully connected layer with 64 hidden units to process this aggregation. Finally, a sigmoid activation is performed to output the predicted probabilities for the input to be either safe or DDoS. The proposed model's architecture is illustrated in Figure 4.

Figure 4. Self-Attention Enabled Weighted Ensemble System Architecture

The proposed model addresses the gaps in existing literature as follows:

1. **Enhanced Generalization in Feature Selection**: The proposed self-attention-enabled DL framework addresses the limitation of traditional feature selection methods by employing multiple deep learning architectures (CNN with XGBoost, LSTM, and Random Forest). Each model extracts features from different perspectives, and the self-attention mechanism dynamically learns the most relevant features across varying datasets. This adaptive feature extraction improves generalization, even in diverse and real-time environments.

2. **Improved Adaptability to Evolving Attack Patterns**: Unlike traditional ML models, the weighted ensemble of attention-enhanced CNN architectures provides adaptability to new attack patterns. The ensemble effectively handles the complexity of modern network environments by integrating self-attention mechanisms and continuously updating the weights based on individual model performance.

3. **Robust Detection of Novel Attacks**: The incorporation of a stacking-based meta-classifier allows the ensemble to combine the strengths of all three models. This approach enhances the model's ability to detect known attack types and novel and unknown threats. The ensemble's capability to adapt to new features extracted from unseen attacks ensures a more robust defense against zero-day attacks, addressing the challenge of handling novel threats.

The model was trained using the Adam optimizer for 64 epochs using the binary cross entropy loss criterion.

$$loss = -\frac{1}{N}\sum_{i=1}^{N}\big(y_i \cdot log\ log\ (p_i)\ + (1 - y_i) \cdot (1 - log\ log\ (p_i)\ )\big) \quad (11)$$

Where,

$N \leftarrow total\ number\ of\ samples\ in\ the\ dataset$

$y_i \leftarrow true\ label\ for\ the\ i^{th}\ sample$

$p_i \leftarrow$
$predicted\ probability\ of\ the\ positive\ class\ for\ the\ i^{th}\ sample$

## IV. EXPERIMENTAL RESULTS

In this section, a thorough analysis is conducted to compare the performance of the proposed model with existing approaches. The models are evaluated using various performance metrics, such as the confusion matrix, accuracy (Equation 12), loss (Equation 11), precision (Equation 13), recall (Equation 14), and F1-score (Equation 15).

$$accuracy \leftarrow \frac{TP + TN}{TP + TN + FP + FN} \quad (12)$$

$$precision \leftarrow \frac{TP}{TP + FP} \quad (13)$$

$$recall \leftarrow \frac{TP}{TP + FN} \quad (14)$$

$$F1\ Score \leftarrow \frac{2 \cdot TP}{2 \cdot TP + TN + FP} \quad (15)$$

Where,
$TP \leftarrow true\ positives$

$TN \leftarrow true\ negatives$

$FP \leftarrow false\ positives$

$FN \leftarrow false\ negatives$

These metrics offer a thorough assessment of the model's effectiveness, highlighting the rigorous analysis and empirical evidence supporting the research. The individual models - SA-Enabled CNN with XGBoost, SA-Enabled CNN with LSTM, and SA-Enabled CNN with Random Forest - were trained separately on the CIC-DDoS dataset, and their results were recorded. Table 2 showcases the performance assessment of these models, serving as the basis for determining each model's contribution weight to the ensemble. Combining the entire predictions from each model is not effective, since a model might be biased towards a particular feature set for certain inputs. Thus, a weighted ensemble method is proposed in this research, which is followed by a fully connected layer to process these final weighted predictions to obtain final outputs. Table 3 comprehensively tabulates the performance comparison among these models, conducted based on the evaluation metrics outlined in Figure 1.

TABLE 2

PERFORMANCE EVALUATION OF SA-ENABLED CNN WITH XGBOOST, SA-ENABLED CNN WITH LSTM, AND SA-ENABLED CNN WITH RANDOM FOREST FOR DETERMINING WEIGHTS FOR THE PROPOSED ENSEMBLE MODEL.

| MODELS | ACCURACY | PRECISION | RECALL | F1-SCORE |
|---|---|---|---|---|
| SA-Enabled CNN with XGBoost (3.2.1) | 0.958 | 0.955 | 0.953 | 0.954 |
| SA-Enabled CNN with LSTM (3.2.2) | 0.971 | 0.973 | 0.971 | 0.972 |
| SA-Enabled CNN with Random Forest (3.2.3) | 0.940 | 0.942 | 0.939 | 0.940 |

During the training process, the model consistently reduces its loss, reaching a final value of 0.005. This decline in loss signifies successful optimization of the model's parameters across the total 64 epochs, resulting in improved convergence towards the desired output. The accuracy curve displays a significant upward trajectory, indicating a continuous improvement in the model's ability to classify DDoS attacks. This steady increase in accuracy underscores the capacity of the model to learn from the training data and generalize patterns, ultimately leading to more precise predictions. The loss and accuracy curves for training and validation sets obtained while training the proposed model have been displayed in Figures 5 and 6 respectively, and the normalized confusion matrix depicting its performance in distinguishing between DDoS and non-DDoS traffic is illustrated in Figure 7.

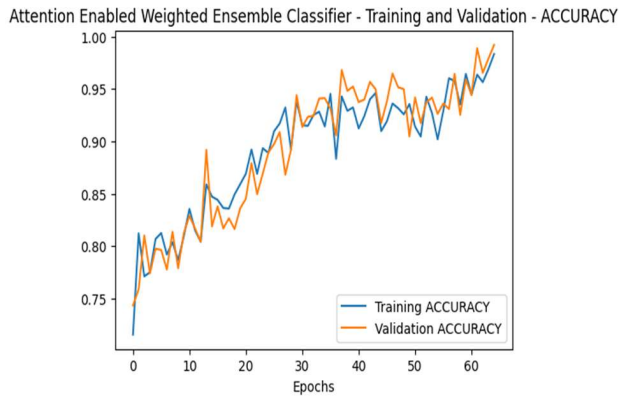

Figure 5. Training And Validation Accuracy Curves Obtained During the Proposed Model Training

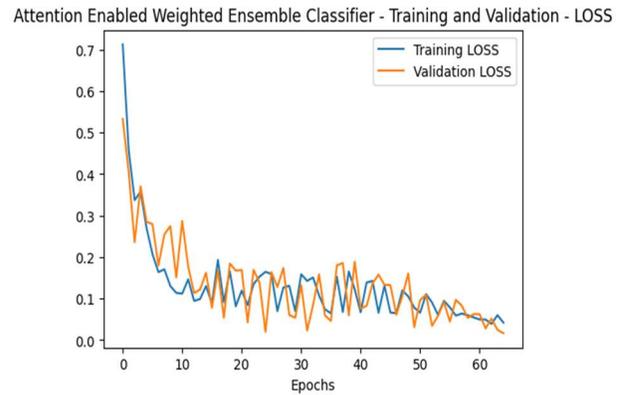

Figure 6. Training And Validation Loss Curves Obtained During the Proposed Model Training.

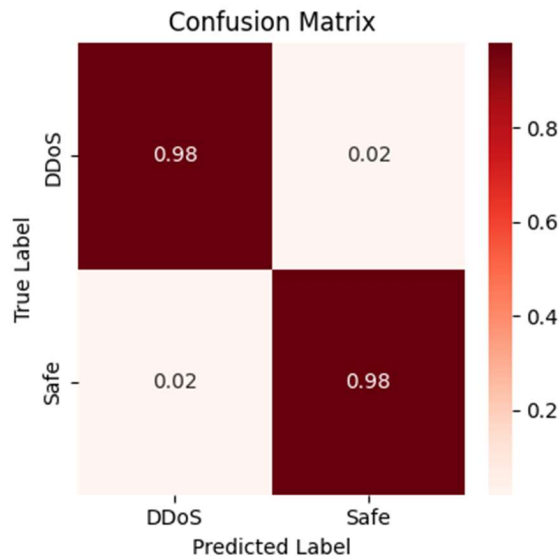

Figure 7. Normalized Confusion Matrix Derived From The Proposed Model

The confusion matrix is a square matrix, where the rows correspond to the actual classes, and the columns represent the predicted classes. It shows the values predicted by the proposed model in relation to the actual values. From Figure 7, it can be seen that the proposed model has correctly classified 98% of the incoming traffic as DDoS or safe.

TABLE 3

PERFORMANCE COMPARISON OF THE PROPOSED MODEL WITH EXISTING MODELS.

| MODELS | ACCURACY | PRECISION | RECALL | F1-SCORE |
|---|---|---|---|---|
| FS + MLP (2021) [17] | 0.9012 | 0.9116 | 0.7941 | 0.7939 |
| AE + Isolation Forest (2020) [18] | 0.8898 | 0.8792 | 0.9348 | 0.9061 |
| Bi-LSTM (2021) [21] | 0.9818 | 0.9793 | 0.9984 | 0.9887 |
| MLP (2020) [23] | 0.9250 | 0.8440 | 0.9420 | 0.8900 |
| D-MLP (2020) [24] | 0.9250 | 0.8340 | 0.9570 | 0.8900 |
| CNN (2020) [24] | 0.9540 | 0.9330 | 0.9240 | 0.9280 |
| Logistic Regression (2020) [24] | 0.8780 | 0.8680 | 0.7710 | 0.7940 |
| Extended Naïve Bayes (2020) [25] | 0.9625 | 0.9600 | 0.9600 | 0.9600 |
| Extended Decision Tree (2020) [25] | 0.8839 | 0.8544 | 0.9595 | 0.9040 |
| AE + Regression (2016) (2021) [26] | 0.9700 | 0.9900 | 0.9700 | 0.9780 |
| **Proposed Model (SA-Enabled Ensemble Classifier)** | **0.9869** | **0.9871** | **0.9863** | **0.9866** |

Table 3 illustrates that the proposed model has outperformed all other compared works. The novel SA-enabled weighted ensemble architecture excelled at feature extraction and combination, ultimately classifying network intrusions as DDoS or safe. The individual CNNs capture a wide range of features, and the SAM at the end enhances these features, strengthening the model's capability to improve its predictions. Upon aggregation using stacking-based weighted ensembling, the final predictions of the proposed model attained an accuracy of 98.69% on the validation data. Bi-LSTM (2021) [21] is the closest competitor, with excellent performance metrics, especially its near-perfect recall (99.84%), making it highly effective in correctly identifying DDoS attacks. However, its slightly lower precision compared to the proposed model suggests a trade-off between identifying actual DDoS attacks and minimizing false positives. AE + Regression (2021) [26] and Extended Naïve Bayes (2020) [25] also show strong results, with the former achieving very high precision (99.00%) and the latter exhibiting balanced performance across all metrics. Despite this, both models fall short in overall performance when compared to the proposed model. Models like MLP (2020) [23] and D-MLP (2020) [24] show decent accuracy but have noticeable discrepancies in precision and recall, which impacts their F1-scores, indicating less consistent performance across different types of network traffic. The CNN (2020) [24] demonstrates relatively high accuracy (95.40%) and balanced precision and recall, making it a solid choice, but it still lags behind the proposed model. On the other hand, the Extended Decision Tree (2020) [25] and AE + Isolation Forest (2020) [18] models present mixed results. While the Extended Decision Tree achieves decent recall, its precision is lower compared to the top-performing models. Similarly, AE + Isolation Forest has high recall but lacks precision. Lastly, FS + MLP (2021) [17] and Logistic Regression (2020) [24] exhibit the least favorable performance, particularly in recall, which suggests that these models are less effective in capturing the full spectrum of DDoS attack characteristics. Overall, the proposed weighted ensemble classifier demonstrates superior performance across all evaluation metrics, outperforming existing research works and establishing itself as a highly effective solution for DDoS attack detection.

## V. CONCLUSION

The research presents a thorough investigation into the application of advanced ML and DL techniques to detect DDoS attacks within the sector of network security and proposes a self-attention-enabled ensemble classifier for the same. It showcases the potential of high-performing DL models such as CNNs and LSTMs, and self-attention mechanisms in combination with ML models such as XGBoost and Random Forest to increase the robustness and performance of DDoS attack detection systems. In essence, it underscores the importance of employing ML and DL algorithms to enhance network security defenses and address evolving cyber threats.

## VI. FUTURE WORK

This research investigated the efficacy of DL models in NID, yielding promising findings. Future work can delve deeper into addressing the vulnerability of the model to adversarial attacks, ensuring robustness against malicious manipulation of input data to evade detection or trigger false alarms. Additionally, efforts can focus on scaling the model to large-scale network environments with high volumes of traffic, overcoming challenges related to computational efficiency and resource management. Furthermore, enhancing the model's ability to generalize effectively to unseen DDoS attack patterns or variations in network conditions will be essential for advancing its practical utility in real-world scenarios.

## VII. STATEMENTS AND DECLARATIONS

**FUNDING:** No funding was received for conducting this research.
**CONFLICTS OF INTERESTS:** The authors do not have any competing interests related to the content of this article to disclose.

**AVAILABILITY OF MATERIAL AND DATA:** Not Applicable.

**CODE AVAILABILITY:** Not Applicable.

## VII. AUTHOR BIOGRAPHY

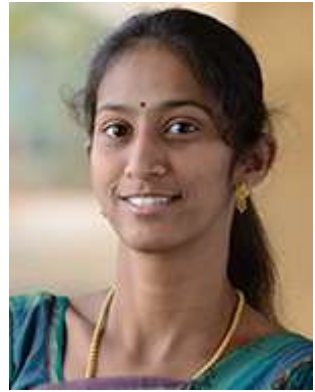

**Dr. Kanthimathi S** is presently associated with Vellore Institute of Technology, Chennai, Tamil Nadu. She earned her Ph.D. in Computer Science and Engineering from Visveswaraya Technological University, Karnataka. She specializes in the research areas of Cloud Computing, Network Security, and Machine Learning

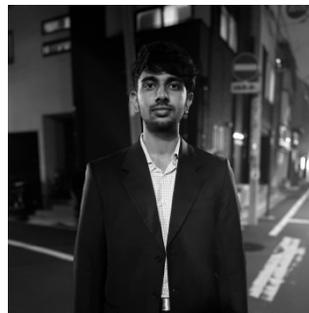

**Shravan Venkatraman**, born in Chennai, India, is a B.Tech Computer Science Engineering student at Vellore Institute of Technology, Chennai. With over two years of deep research experience, Shravan has delved into artificial intelligence, machine learning, deep learning, computer vision, generative AI, and software development. As an AI engineer, he is passionate about streamlining innovative solutions for real-world challenges. His academic journey is marked by multiple presentations at international conferences and publications in high-impact journals, demonstrating his commitment to advancing technology. Shravan's dedication to continuous learning and problem-solving is evident in his active participation in research and development, where he persistently seeks new frontiers to explore and conquer.

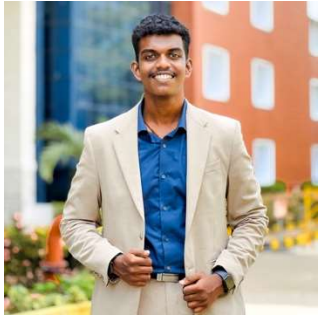

**Jayasankar K S**, born in Kerala, India is a B.Tech Computer Science Engineering student at Vellore Institute of Technology, Chennai. With over two years of research expertise, he has played an active role in various projects focusing on cybersecurity, artificial intelligence and machine learning. Furthermore, he has led a team to victory at an international UI/UX competition showcasing his expertise in the field of Human-Computer Interaction.

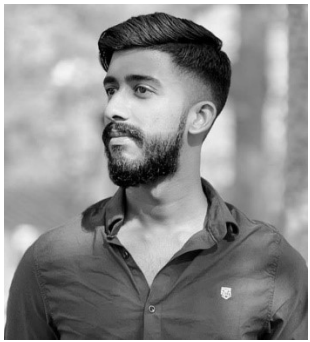

**Pranay Jiljith**, born in Kerala, India, is a B.Tech Computer Science Engineering student at Vellore Institute of Technology, Chennai. With over two years of expansive research experience, he has done various studies, researches and honed skills in cybersecurity, IoT, and emerging technology. A content creator and tech influencer, Pranay's expertise is reflected in his victory at an international UI/UX competition. He also has a passion for filmmaking, blending creativity with his technical knowledge. Actively involved in various events and research, Pranay is committed to advancing in the tech field while continuously exploring new challenges.

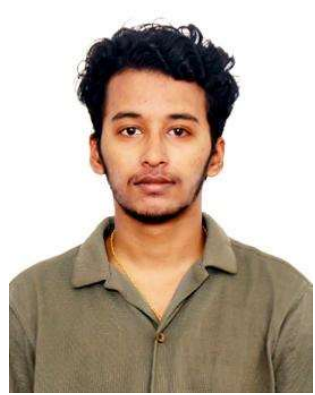

**Jashwanth R** is a dedicated Computer Science and Engineering student at VIT Chennai with a strong focus on web development, UI/UX design, and machine learning. Over the years, he has honed his skills in creating innovative and user-friendly digital experiences, contributing to a range of projects that showcase his expertise in the development domain. In addition to his work in web development, Jashwanth has also completed and published projects in machine learning, demonstrating his versatility and commitment to staying at the forefront of technology.